\documentclass[conference]{IEEEtran}
\usepackage[font=small,labelfont=bf]{caption}
\usepackage{graphicx, balance}
\usepackage{caption,subcaption}
\usepackage{linguex}
\usepackage[linguistics,edges]{forest}
\usepackage{enumitem}
\usepackage{booktabs}
\usepackage{multirow, makecell}
\usepackage[cmex10]{amsmath}
\usepackage{algorithmic}
\usepackage{url}
\usepackage{csquotes} 
\usepackage{arydshln} 
\usepackage{color, xcolor, soul}
\usepackage{amssymb}
\usepackage{makecell}
\usepackage{cite}
\usepackage{etoolbox}
\usepackage{threeparttable}
\usepackage{hyperref}
\usepackage{pifont}

\usepackage{array}
\usepackage{booktabs}
\usepackage{multirow}
\usepackage{tabularx}

\usepackage{rotating}

\definecolor{StartStopColor}{HTML}{FC427B} 
\definecolor{InputOutputColor}{HTML}{25CCF7} 
\definecolor{ProcessColor}{HTML}{29F76E} 
\definecolor{DecisionColor}{HTML}{A7257D}
\definecolor{textcolor}{HTML}{000000}
\definecolor{processtextcolor}{HTML}{DDEA40}

\usepackage{tikz}
\usetikzlibrary{shapes.geometric, arrows}
\tikzstyle{startstop} = [rectangle, rounded corners, minimum width=3cm, minimum height=1cm, text centered, draw=black, text=textcolor, fill=StartStopColor!70]
\tikzstyle{io} = [trapezium, trapezium left angle=70, trapezium right angle=110, text width=5cm, minimum height=1cm, text centered, draw=black, fill=InputOutputColor!70, text=textcolor]
\tikzstyle{process} = [rectangle, text width=8cm, minimum height=1cm, text centered, draw=black, fill=ProcessColor!70, text=textcolor]
\tikzstyle{decision} = [diamond, minimum width=3cm, minimum height=1cm, text centered, draw=black, fill=green!30]
\tikzstyle{arrow} = [thick,->,>=stealth]

\title{Mitigating Bias in Machine Learning Models \\ for Phishing Webpage Detection }
\author{\IEEEauthorblockN{Aditya Kulkarni\textsuperscript{*}, Vivek Balachandran\textsuperscript{\#}, Dinil Mon Divakaran\textsuperscript{\$} and Tamal Das\textsuperscript{*}}
\IEEEauthorblockA{\textsuperscript{*}\emph{Indian Institute of Technology, Dharwad, India} \\
\textsuperscript{\#}\emph{Singapore Institute of Technology, Singapore} \\
\textsuperscript{\$}\emph{National University of Singapore, Singapore} \\
\textsuperscript{*}\{aditya.kulkarni, tamal\}@iitdh.ac.in, \textsuperscript{\#}vivek.b@singaporetech.edu.sg, \textsuperscript{\$}dinil@comp.nus.edu.sg}}

\begin{document}

\maketitle

\begin{abstract}
    The widespread accessibility of the Internet has led to a surge in online fraudulent activities, underscoring the necessity of shielding users' sensitive information from cybercriminals. \textit{Phishing}, a well-known cyberattack, revolves around the creation of phishing webpages and the dissemination of corresponding URLs, aiming to deceive users into sharing their sensitive information, often for identity theft or financial gain. Various techniques are available for preemptively categorizing \textit{zero-day} phishing URLs by distilling unique attributes and constructing predictive models. However, these existing techniques encounter unresolved issues. This proposal delves into persistent challenges within phishing detection solutions, particularly concentrated on the preliminary phase of assembling comprehensive datasets, and proposes a potential solution in the form of a tool engineered to alleviate bias in ML models. Such a tool can generate phishing webpages for any given set of legitimate URLs, infusing randomly selected content and visual-based phishing features. Furthermore, we contend that the tool holds the potential to assess the efficacy of existing phishing detection solutions, especially those trained on confined datasets.
\end{abstract}


\section{Introduction}
\label{sec:Introduction}
In today's digital age, the increasing reliance of users on online services, such as e-commerce, online banking, and social media platforms, has become exceedingly prominent. These virtual platforms offer unparalleled convenience, allowing users to access services effortlessly from any location, greatly simplifying their lives. However, the utilization of these digital services necessitates the submission of sensitive user information, such as credit/debit card particulars and login credentials. This is essential for delivering personalized and precise services, streamlining the user experience. Unfortunately, the very convenience that these online services offer also presents a vulnerability, providing malicious actors with an opportunity to exploit and collect users' sensitive data, thereby instigating cyberattacks.

\begin{figure}[!t]
    \centering
    \includegraphics[width=0.49\textwidth, height=0.293\textwidth]{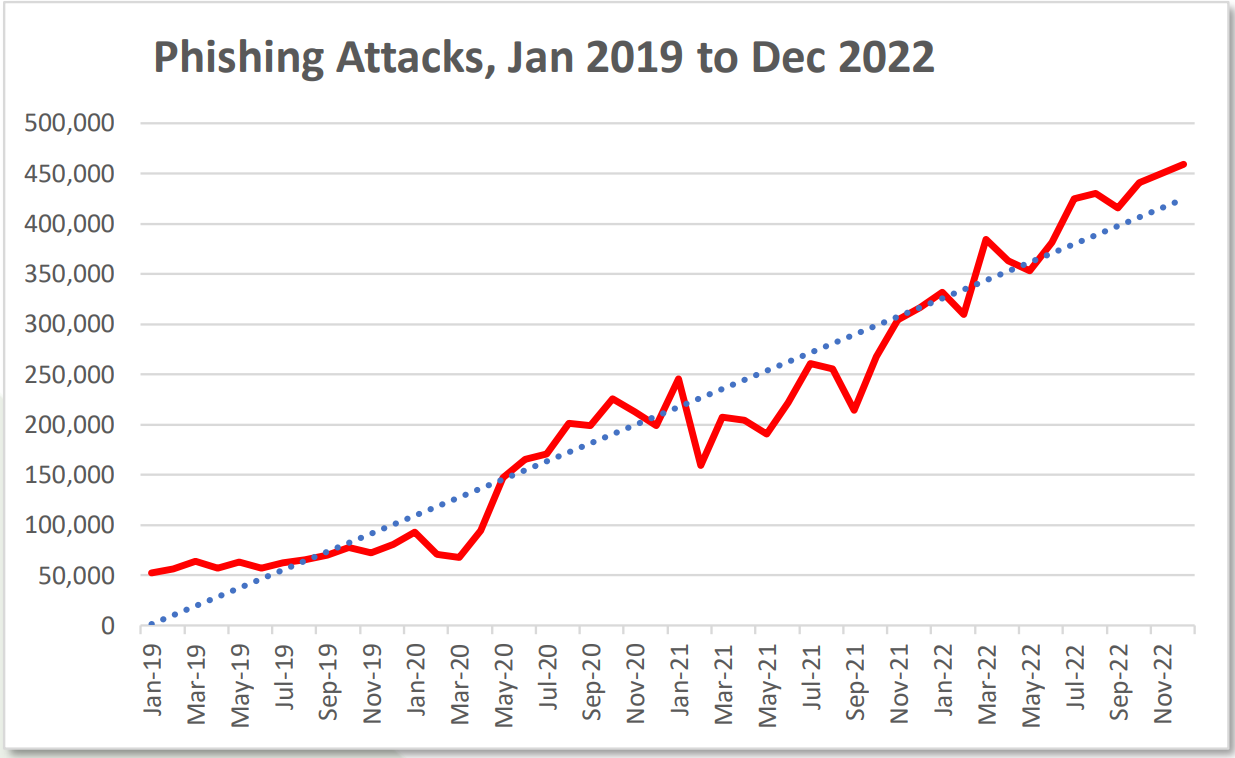}
    \caption{\centering Phishing Attacks from Jan $2019$ to Dec $2022$: APWG~\cite{APWG_REPORT_4_2022}}
    \label{fig:APWG_Report_4_Years}
\end{figure}

A spectrum of cyberattacks revolves around the objective of amassing users' confidential data, which can then be utilized for malicious purposes or financial gains. One prominent example of such malevolent activities is \textit{phishing}. Phishing attacks are initiated by creating a phishing webpage that mimics a legitimate webpage. This phishing webpage, infused with malicious features, is formulated to deceive users into gathering their sensitive information. The created phishing webpage is subsequently hosted via a phishing URL, which is disseminated to potential victims via malicious emails, SMS, social media messages, and similar channels. Upon clicking the phishing URL, users are redirected to the corresponding phishing webpage and are prompted to input their sensitive information into the input fields provided. Their sensitive information is either dispatched to the attackers' email or incorporated into their database. The ill-intentioned individuals behind the attack then exploit the pilfered sensitive information, engaging in identity theft or leveraging it for financial gains. 
Figure~\ref{fig:APWG_Report_4_Years} depicts a notable upsurge in worldwide phishing attacks over the past four years, accompanied by a progressively steep rise of $150\%$ annually.

\section{Related Work}
\label{sec:Taxonomy_of_Phishing_Detection_Approaches}
In response to this escalating threat, researchers have endeavoured to devise various defensive strategies to safeguard users' sensitive data. A multitude of techniques have been proposed for the detection of phishing attempts. These techniques entail scrutinizing the phishing URLs, assessing the content analyzing the visual composition of the phishing webpages. Through meticulous examination, these techniques classify the webpage into one of two categories: \textit{phishing} or \textit{legitimate}. This ongoing effort to enhance security mechanisms safeguards users against phishing attacks.

URL-based approaches for phishing detection include techniques based on \textit{lists}, \textit{heuristics}, and \textit{machine learning (ML)}. The conventional list-based approach employs pre-defined \textit{whitelists} and \textit{blacklists} to classify input URLs as legitimate or phishing, respectively~\cite{cao2008anti}. However, it is ineffective against \emph{zero-day attacks} -- newly created phishing websites which are yet to be classified.

To address the limitations of list-based techniques, alternative strategies such as heuristics and ML-based techniques have been developed. These methods examine distinctive characteristics present in both legitimate and phishing webpages. When encountering a suspicious URL, relevant attributes are extracted and compared against recognized phishing patterns. If the attributes align with established phishing indicators, the URL is identified as phishing, overcoming the challenges posed by zero-day attacks~\cite{ma2009beyond, zhang2007cantina}

The ML-driven approaches entail gathering URLs and webpage contents from repositories such as PhishTank\footnote{PhishTank: \url{https://phishtank.org/phish_archive.php}}, UC Irvine\footnote{UC Irvine\url{https://archive.ics.uci.edu/ml/datasets/phishing+websites}}. This is followed by \textit{selecting} features capable of distinguishing between phishing and legitimate inputs. Once the feature selection is completed, the relevant features are extracted to create a feature vector that is taken as input to train a range of classifiers, including SVM, AdaBoost, Naive Bayes, Random Forest, and Decision Tree~\cite{shirazi2018kn0w}. The trained classifiers can then accurately classify zero-day phishing URLs. However, imbalanced datasets with more legitimate (than phishing) samples (or vice versa) and limited diversity introduce classification bias.

The visual analysis for distinguishing between phishing and legitimate webpages incorporates two distinct techniques: \textit{webpage screenshot similarity} and \textit{DL-based} approach. In the webpage screenshot technique~\cite{afroz2011phishzoo}, an archive housing screenshots of legitimate webpages, alongside their corresponding domain names, is maintained. When presented with a suspicious URL, the corresponding webpage is accessed to generate a screenshot. This screenshot is then compared to those stored in the database, resulting in a similarity score. A high similarity score triggers a comparison of domain names. If both the webpage screenshot and the domain name match, the suspicious URL is classified as legitimate. Conversely, if the webpage screenshot shows a high similarity score, but the domain name doesn't match, the suspicious URL is labelled as phishing.

Furthermore, utilizing a DL-based approach enhances the process of extracting features used in categorizing a suspicious URL as phishing or legitimate. This approach employs neural networks for detection, leading to the attainment of high accuracy in prediction. VisualPhishNet~\cite{abdelnabi2020visualphishnet} trained a Siamese model using screenshots from secure websites to detect visual similarities with well-established legitimate websites. Some studies adopt a hybrid approach that combines heuristic-based features of URL and webpage attributes to train ML, and DL algorithms~\cite{somesha2020efficient} to enhance phishing detection accuracy.

\section{Problem Statement}
\label{sec:Open_Issues_and_Possible_Solutions}
Drawing upon extensive research on diverse phishing detection methodologies documented in the literature, we identify the following open issues concerning the pre-processing phase of training ML models.

\begin{description}
    \item \textbf{Biased datasets:} Classifiers learn distinctive features from training data to construct \emph{decision trees} or \emph{classification rules}~\cite{freitas2014comprehensible}. Using imbalanced datasets (i.e. a disproportionate number of phishing and legitimate URLs) in phishing detection can introduce bias in classifiers' tree or rule-building process. Consequently, the model skews towards the larger dataset, reducing accuracy in classifying new suspicious URLs.

    \item \textbf{Diverse dataset:} Researchers gather phishing samples from various repositories to create a diverse dataset. Sourcing from a single repository may limit diverse phishing characteristics, causing misclassification by ML classifiers. A diverse dataset is necessary for accurate phishing detection, including URL and webpage content features (logos, favicons, HTML, CSS, and JS/PHP codes).
\end{description}

ML models, when trained on such confined datasets (characterized by imbalanced samples and limited feature diversity), tend to exhibit a bias, consequently diminishing their accuracy in effectively identifying zero-day phishing attacks. We aim to create a tool to mitigate these challenges.

\section{Proposed Solution}
\label{sec:Potential_Solutions}

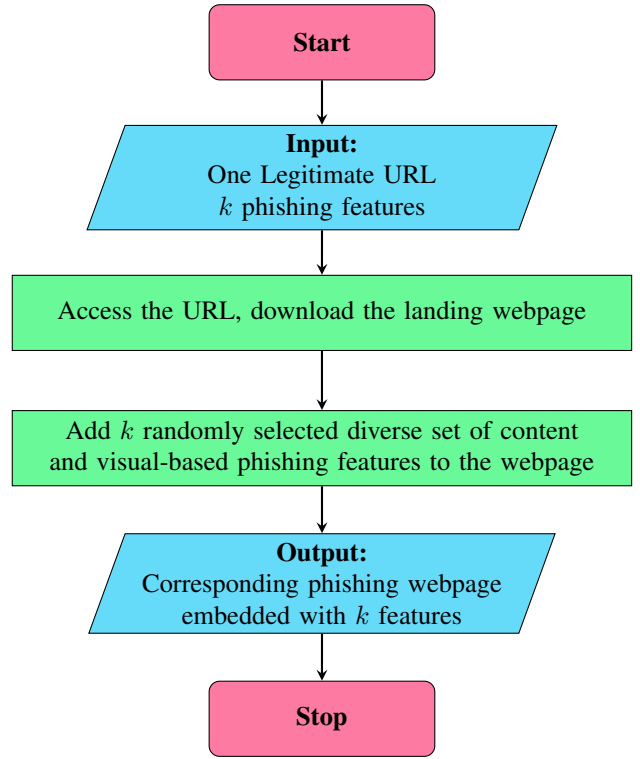
\begin{figure}[th]
    \centering
    \begin{tikzpicture}[node distance=1.8cm]
        \node (start) [startstop] {\textbf{Start}};
        \node (in) [io, below of=start] {\shortstack{\textbf{Input:} \\ One Legitimate URL \\ $k$ phishing features}};
        \node (pro1) [process, below of=in] {\shortstack{Access the URL, download the landing webpage}};
        \node (pro2) [process, below of=pro1] {\shortstack{Add $k$ randomly selected diverse set of content \\ and visual-based phishing features to the webpage}};
        \node (out) [io, below of=pro2] {\shortstack{\textbf{Output:} \\ Corresponding phishing webpage \\ embedded with $k$ features}};
        \node (stop) [startstop, below of=out] {\textbf{Stop}};
    
        \draw [arrow] (start) -- (in);
        \draw [arrow] (in) -- (pro1);
        \draw [arrow] (pro1) -- (pro2);
        \draw [arrow] (pro2) -- (out);
        \draw [arrow] (out) -- (stop);
    \end{tikzpicture}
\caption{Phishing Webpage Generation Tool}
\label{fig:tool_flowchart}
\end{figure}

\noindent
To address the difficulties arising from confined datasets (which encompass biased datasets and a lack of diversity), we're in the process of developing a solution for generating phishing webpages. This automated tool functions by taking a legitimate URL, extracting its source code and incorporating random content and visual-based phishing features to generate corresponding phishing webpage. The process flow of this tool for generating phishing webpages is illustrated in Figure~\ref{fig:tool_flowchart}. The dataset produced by utilizing this tool maintains a balance between legitimate and phishing webpages, thereby ensuring a well-rounded and varied dataset due to the random inclusion of phishing features. These phishing webpages can then be utilized to evaluate the efficacy of existing ML-based phishing detection solutions that generally rely on confined datasets.

\begin{figure}[h]
    \centering
    \includegraphics[width=0.49\textwidth,]{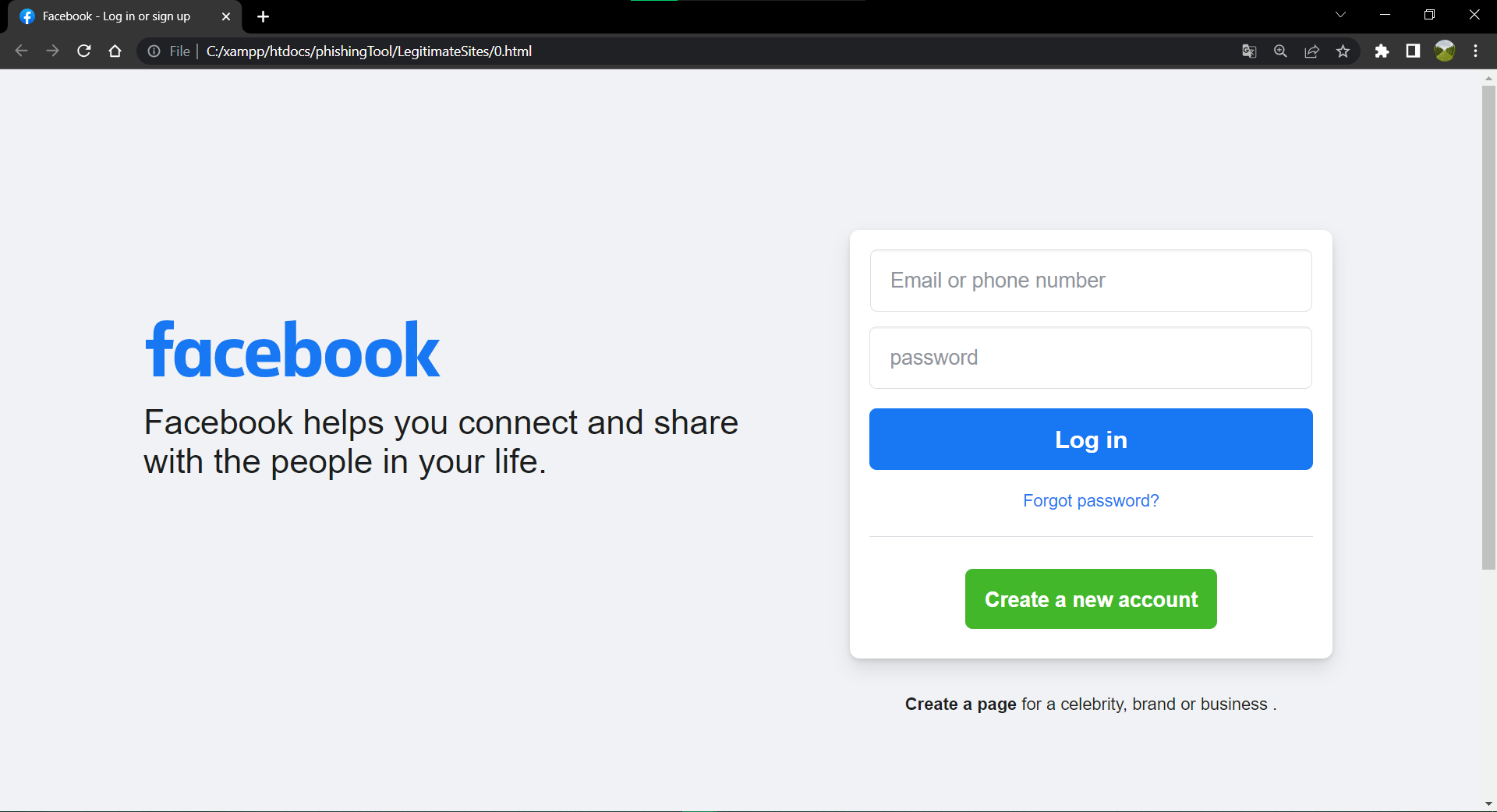}
    \caption{Legitimate Webpage}
    \label{fig:legitimate_webpage}
\end{figure}
\vspace{-0.25cm}
\begin{figure}[h]
    \centering
    \includegraphics[width=0.49\textwidth,]{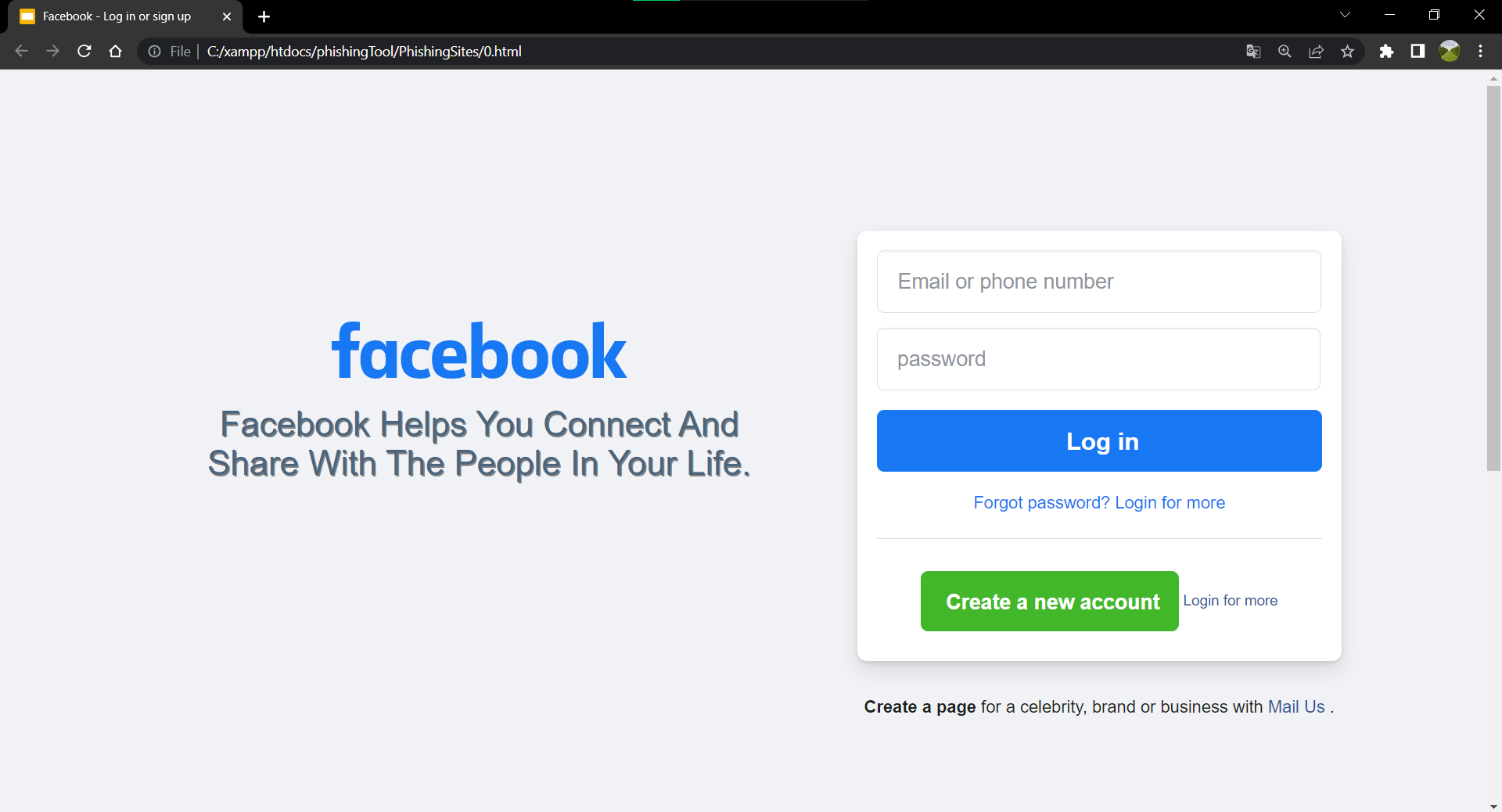}
    \caption{Generated Phishing Webpage}
    \label{fig:generated_phishing_webpage}
\end{figure}

Our automated tool's characteristics in the context of phishing involve adaptability in generating phishing webpages using legitimate URLs. It achieves this dynamism by generating phishing webpages to match legitimate ones, coupled with diversity through the integration of randomized content and visual-based phishing features. For instance, a genuine webpage incorporating a \texttt{<form action="legitimate\_source">} element will undergo alteration, substituting the \texttt{action=""} value with a malicious counterpart. This modification aims to redirect sensitive user data to the attacker. Another deception involves the manipulation of anchor tags (\texttt{<a>}) within legitimate webpage content. These tags, originally linked to lawful destinations, will be exchanged with \texttt{href} attributes like \texttt{"\#"} or \texttt{"\#content"} or \texttt{"Javascript:void(0)"} or a malicious link. Additionally, a suite of visual-based phishing features will alter the rendering of legitimate webpages. Alterations may include changing the appearance of favicons and logos. Genuine favicons and logos present on lawful pages will be substituted with lighter or darker iterations, or even replaced entirely by alternate favicons and resemblant logos. For a given legitimate webpage in Figure~\ref{fig:legitimate_webpage}, the tool adds randomly selected content-based phishing features, such as a malicious \texttt{<form>} tag, modifications in the \texttt{<a href="">} tag, and several distinct visual-based phishing features encompassing font stylization, alternate favicon usage, and opacity adjustments for the webpage. Subsequently, this process results in the creation of an associated phishing webpage, as depicted in Figure~\ref{fig:generated_phishing_webpage}.

\section{Conclusion and Future Scope}
\label{sec:Conclusion_and_Future_Work}
The ever-increasing threat of phishing, a well-recognized cyberattack, capitalizes on phishing URLs to manipulate users into divulging confidential information, particularly targeting financial websites where intricately fabricated counterfeit versions aim to capture users' sensitive information. List-based techniques prove inadequate against zero-day phishing URLs which are addressed by heuristic, ML and DL-based approaches. This study has delved into the persistent challenges inherent in phishing detection solutions, focusing particularly on the foundational phase of dataset compilation. In response, we have introduced a prospective solution in the form of a tool designed to address dataset-related issues. By generating phishing webpages for a given set of legitimate URLs, incorporating randomly selected content and visual-based phishing features, our tool offers a potential path forward. Furthermore, we assert that our tool can evaluate the effectiveness of existing phishing detection solutions, especially those trained on confined datasets.

\bibliographystyle{IEEEtran}
\bibliography{references}

\end{document}